\documentclass[manuscript]{acmart}
\AtBeginDocument{%
  \providecommand\BibTeX{{%
    \normalfont B\kern-0.5em{\scshape i\kern-0.25em b}\kern-0.8em\TeX}}}

\setcopyright{acmcopyright}
\begin{document}
\acmConference{}{}{}
\acmYear{}
\copyrightyear{}
\acmISBN{}
\acmDOI{}
\acmPrice{}
\title{Sentiment Analysis in Learning Management Systems Understanding Student Feedback at Scale}

\author{Mohammed Almutairi}
\affiliation{%
  \institution{University of Notre Dame}
  \city{Notre Dame}
  \country{USA}}
\email{malmutai@nd.edu}








\renewcommand{\shortauthors}{Almutairi}

\begin{abstract}
  During the wake of the Covid-19 pandemic, the educational paradigm has experienced a major change from in person learning traditional to online platforms. The change of learning convention has impacted the teacher - student especially in non-verbal communication. The absent of non-verbal communication has led to a reliance on verbal feedback which diminished the efficacy of the educational experience. This paper explores the integration of sentiment analysis into learning management systems (LMS) to bridge the student-teacher's gap by offering an alternative approach to interpreting student feedback beyond its verbal context. The research involves data preparation, feature selection, and the development of a deep neural network model encompassing word embedding, LSTM, and attention mechanisms. This model is compared against a logistic regression baseline to evaluate its efficacy in understanding student feedback. The study aims to bridge the communication gap between instructors and students in online learning environments, offering insights into the emotional context of student feedback and ultimately improving the quality of online education. 
\end{abstract}


\begin{teaserfigure}
  \includegraphics[width=\textwidth]{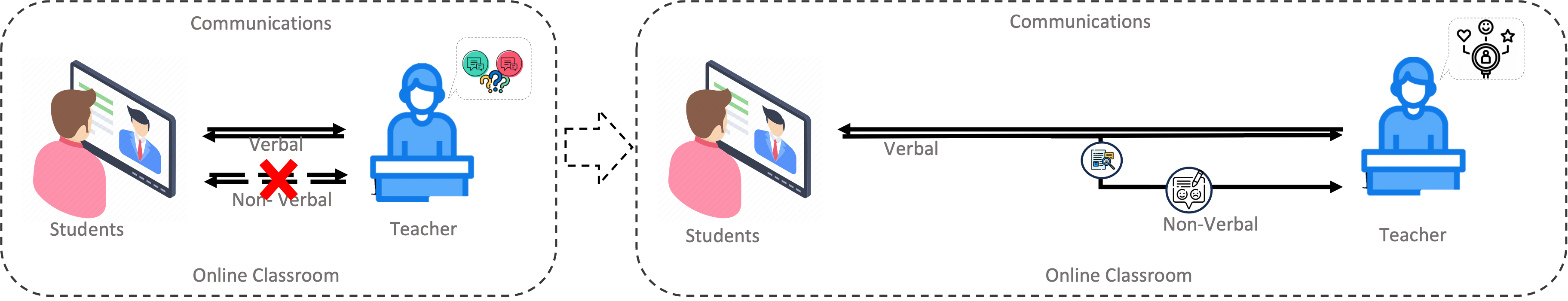}
  \caption{Bridging the Non-Verbal Communication Gap in Online Education through Sentiment Analysis Integration in Learning Management Systems}
  \label{fig:teaser}
\end{teaserfigure}



\maketitle

\section{Introduction}
The education industry confronted unprecedented challenges during the spread of Covid-19, compelling millions of teachers and students to shift to online method of learning~\cite{1}. In 2021, McKinsey highlighted that the proliferation of Covid-19 significantly accelerated the development of online educational platforms and courses~\cite{2}. Yet, it remains largely unexplored whether these newly developed platforms and courses truly cater to the optimal learning and interactions need of students.

Both verbal and non-verbal interactions between students and instructors play an important role in enhancing the learning experience~\cite{3}. In traditional, in-person learning methods, instructors can leverage non-verbal indications, which often significantly influence teaching methods. For instance, instructors can perceive cues like eye contact that mirror confusion or misunderstanding through non-verbal communication. Expressions like narrowed eyes may indicate frustration, while wide-eyed surprise can suggest unexpected realizations~\cite{4}. However, the transition to online learning methods has disrupted communications mode of interaction, specially sidelining non-verbal communication. Consequently, the educational journey has become predominantly reliant on students' verbal feedback. The missingness of non-verbal interactions has expanded the gap between instructors and students, affecting the overall quality of the educational experience~\cite{5}.

While sentiment analysis and opinion mining assist in monitoring feelings about learning contents from student feedback, and providing a deeper understanding beyond verbal communication~\cite{6}, there is existing gap in the literature regrading sentiment analysis during the learning feedback. Various studies have emphasized the importance of considering emotions expressed in feedback's to better diagnose and identify non-verbal expressions ~\cite{11, 12, 13}. However, the exploration sentiment analysis in an education context remain underexplored.

Despite the promising aspects of integrating sentiment analysis into Learning Management Systems (LMS), it remains unknown how learning experiences and educational efficacy will evolve to enhance student experiences. Using sentiment analysis during the learning journey provides a pathway to discern whether feelings and emotions in verbal communications are positive or negative. The implementation of this feature in the LMS can assist with understanding when we cannot observe face-to-face actions between instructors and students~\cite{7}. This paper introduces a sentiment analysis model to be integrated into an LMS. The integration of sentiment analysis functionality aims to enhance the overall learning experience and support students by incorporating their feedback to refine and improve course content, thereby addressing the identified gap in the existing literature.

In addressing the challenges of effectively interpreting student feedback in online learning environments, our initial step involved meticulous data preparation and preprocessing. This crucial phase ensured the consistency and integrity of our dataset, including the resolution of any missing data issues. We then embarked on a strategic process of featurization, where we carefully selected relevant features from the data and created new ones to enhance our analysis. Following this, we selected a logistic regression model as our baseline. This model choice was to establish a foundational benchmark against our model. we developed a sentiment analysis model utilizing a deep neural network. This model is composed of four integral components: a word embedding layer, Long Short-Term Memory (LSTM), an attention mechanism, and an output layer. The final stages of our work focused on the training, testing, and analysing of our model, ensuring its accuracy and efficacy in interpreting student feedback.

\section{Method}
\subsection{Dataset}
We utilized the RateMyProfessor dataset, which captures students' comments and ratings for educational courses. This dataset includes a 'star\_rating' feature that provides an overall student rating on a scale from 1 to 5. These ratings help classify student comments as positive, negative, or neutral. The 'comments' feature contains valuable feedback about courses, teachers, and the educational environment. The RateMyProfessor dataset serves our purpose for several reasons: it offers a diverse range of student feedback on instructors and courses, thereby capturing various institutions and teaching styles. This is something not possible with a single LMS dataset alone. Additionally, This feedback can serve as a proxy for understanding broader student experiences in the academic environment. The dataset's public availability also makes it a more accessible resource compared to many LMS platforms, where feedback is often restricted and private.

\begin{table}
\centering
\caption{Attributes in RateMyProfessor Dataset ~\cite{8}}
\begin{tabular}{|p{0.25\textwidth}|p{0.7\textwidth}|}
\hline
\textbf{Attribute} & \textbf{Description} \\
\hline
Professor name & Name of the professor who is rated. \\
\hline
School name & University where the professor is currently teaching. \\
\hline
Department name & Department where the professor is currently working. \\
\hline
Local name & University's locally known name. \\
\hline
State name & State in which the university is located. \\
\hline
Year since first review & The professor's teaching age, from the first student evaluation to the time when analysis was done in 2019. \\
\hline
Star rating & The star rating of the professor's overall quality. Points 3.5-5.0 are considered good, 2.5-3.4 is average, and 1.0-2.4 is poor according to RMP's official standard. This star rating is the average score given to professors by all student comments. \\
\hline
Take again & Percentage of students who would choose this course again. \\
\hline
Difficulty index & Difficulty level of a course. Point 1 is easiest, and point 5 is hardest. The difficulty index is the average score given to professors by all students. \\
\hline
Tags & Tags that students chose to describe a professor. \\
\hline
Post date & Date when the student posted a course evaluation. \\
\hline
Student star & Star rating given to a professor by each student. \\
\hline
Student-rated difficulty & Difficulty index given to a professor by each student. \\
\hline
Attendance & Whether attendance for a course is mandatory or not. \\
\hline
For credit & Whether students chose a course for credit (yes or no). \\
\hline
Would take again & Whether students would choose a course again (yes or no). \\
\hline
Grade & Student's final score of a course. Possible grades are A+, A, A-, B+, B, B-, C+, C, C-, D+, D, D-, F, WD, INC, Not, Audit/No. "WD" is Drop/Withdrawal, "INC" means Incomplete, "Not" is Not sure yet, and "Audit/No" is Audit/No Grade. \\
\hline
Comment & Comments that students gave for professors. \\
\hline
\end{tabular}
\end{table}

\subsection{Feature extraction}
A set of features, including 'star\_rating', 'diff\_index', 'students\_star', 'student\_difficult', and 'comments', was selected to facilitate necessary data preprocessing. New features, such as 'sentiment', were introduced to represent comment classifications (Positive, Neutral, and Negative) and to serve as independent features for our model.

We transformed text into vector space, utilizing the Term Frequency - Inverse Document Frequency (TF-IDF) method to convert text data into a numerical format. It's important to consider the most occurring words in the dataset, as it enables the model to distinguish words that are indicative of sentiment, thereby enhancing its sensitivity to them. We aim to prevent the model from focusing on the most repeated words in the dataset, especially those not related to sentiment.

The idea of the TF-IDF approach is not to select the most appearing word in student comments but to provide context to them. Furthermore, the preprocessed dataset will be utilized to perform a summary of student comments within a single course. Thus, the summary should focus not on selecting the most appearing words, but on identifying the topic of the comments. Employing TF-IDF is an optimal approach as we seek to elevate the importance of terms appearing exclusively in one class (negative or positive), since such terms could assist the model in class identification, while aiming to minimize the impact of general terms that appear in all classes.

Engaging in sentiment analysis implies that not all vocabulary $\sum$ are of equal importance to our objective. Therefore, we employed a statistical method, the chi-squared test, to measure the relationship between each feature in our vocabulary and the output sentiment class. Features with high chi-squared statistics are considered imperative to our model. 

The vocabulary $\sum$ consists of 17,583 unique words, and the 'comment' feature contains a total of 19,993 sentences. The chi-squared statistics suggest a steady impact when considering a vocabulary of 5000 words, as per the chi-square scores.

\begin{figure}[H]
    \centering
    \begin{minipage}[t]{0.48\textwidth}
        \centering
        \includegraphics[width=\textwidth]{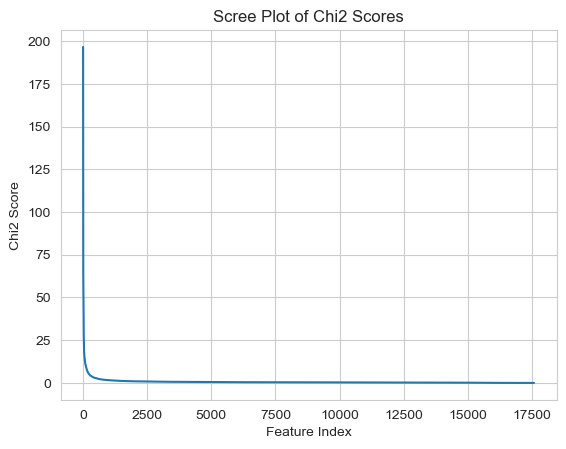}
        \caption{Chi-square scores for each feature, utilized to select the most relevant vocabulary $\sum$ for the sentiment analysis model.}
        \label{fig:chi_square}
    \end{minipage}
    \hfill
    \begin{minipage}[t]{0.48\textwidth}
        \centering
        \includegraphics[width=\textwidth, height=6cm]{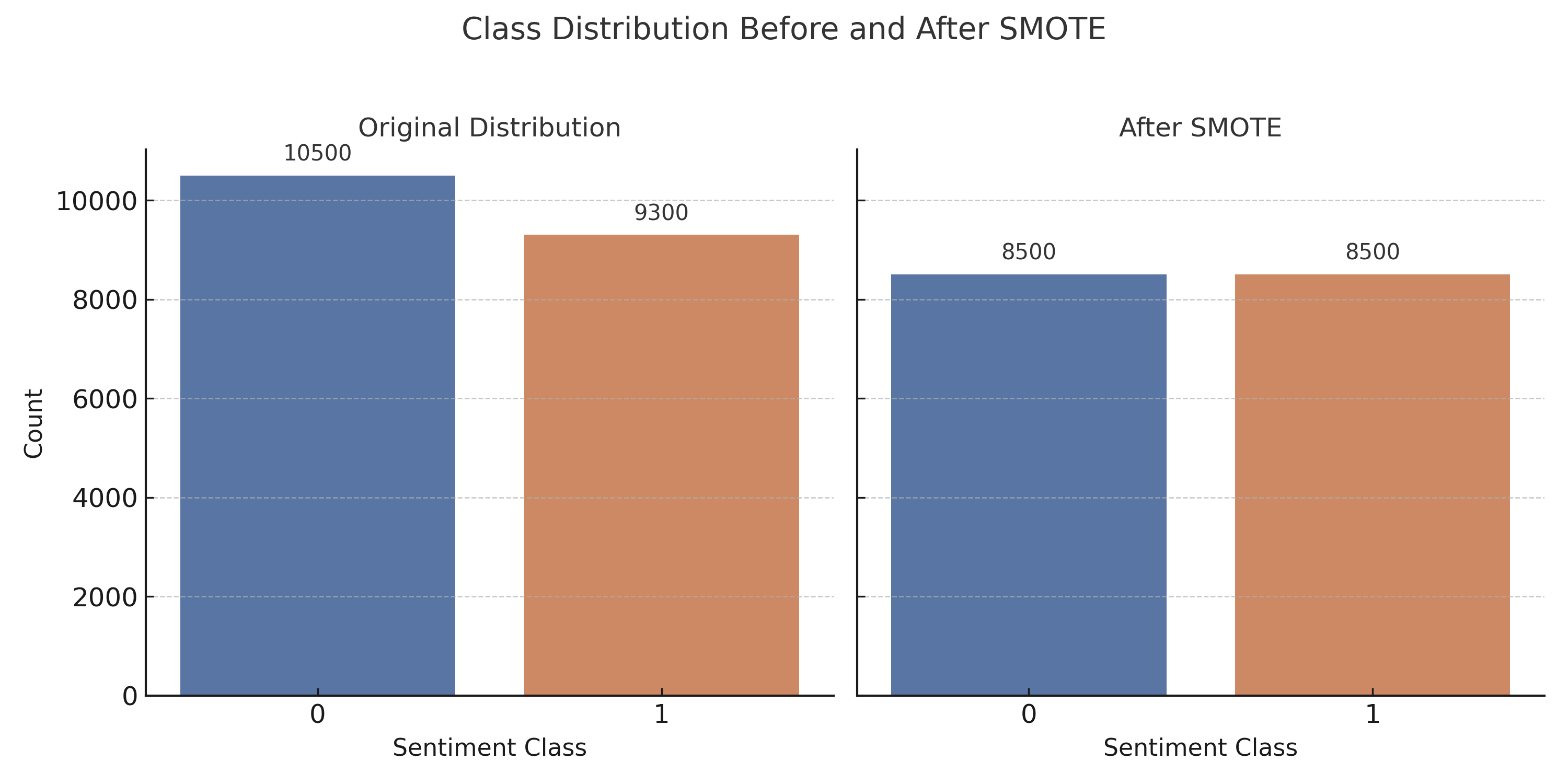}
        \caption{Distribution of the dataset’s classes before and after applying SMOTE to balance the dataset.}
        \label{fig:smote_dist}
    \end{minipage}
\end{figure}

\subsection{Data Preparation and Preprocessing}
The data preparation and preprocessing phase was important in our experimental process, presenting initial challenges related to missing data within our dataset; for example, seven records in the 'comments' feature were completely absent without any systematic pattern. To handle these instances, we opted to drop those records since no appropriate imputation method was applicable in these cases.

We also removed stopwords from the 'comments' feature, such as "the", "an", "in", etc. These stopwords do not contribute meaningful information to the comments, and excluding them improve the model's performance. This was achieved by utilizing the NLTK library~\cite{9}.

We also performed a lemmatization process on the 'comments' feature. Lemmatization assists in grouping together different forms of the same word. For example, some verbs in the comments are in the present progressive form, such as "professor is grading" and "the class is extremely interesting". After performing lemmatization, we can reduce vocabulary dimensionality and help to increase the model's accuracy in sentiment analysis. We utilized the `WordNetLemmatizer` function from the NLTK library ~\cite{10} to perform the lemmatization.

\subsection{Imbalance dataset}

Addressing the challenge posed by the imbalance in the dataset is crucial. We utilized the Synthetic Minority Over-sampling Technique (SMOTE) method to oversample the minority class, as model bias can emerge when dealing with imbalanced datasets, negatively impacting the model's performance. An imbalanced dataset can also diminish the recall, F1 score, and Area Under the Curve (AUC). To prevent this, it's important to ensure that the dataset is balanced. We visualized the dataset distribution to observe the number of comments that are positive and negative, using the star rating as an indicator of dataset distribution. When a class is undersampled, we employ the SMOTE to mitigate imbalance.


\subsection{Model Training and Testing}
To evaluate the performance of our model and the baseline model, we divided the dataset into two subsets: a training set and a test set. The training set, comprising 80\% of the entire dataset, is utilized to train both the baseline and our models. Meanwhile, the test set, making up 20\% of our dataset, is used to validate the generalizability and performance of our model.

\subsection{Baseline model}
We employed a Logistic Regression model as a baseline to establish a fundamental performance benchmark against our models. Logistic Regression was selected due to its simplicity and efficacy as a starting point for binary classification tasks. This model was applied to the dataset following comprehensive pre-processing and feature engineering steps, ensuring that the textual data was transformed into a format suitable for the model learning applications. Subsequent to the model training, it was evaluated utilizing a set of metrics, which will also be used to assess our model, providing a consistent basis for comparison. These metrics and their implications are elaborated upon in Section ~\ref{sec:metrics}.

\subsection{Deep Learning Model}
\begin{figure}[h]

    \centering
    \includegraphics[width=0.8\textwidth]{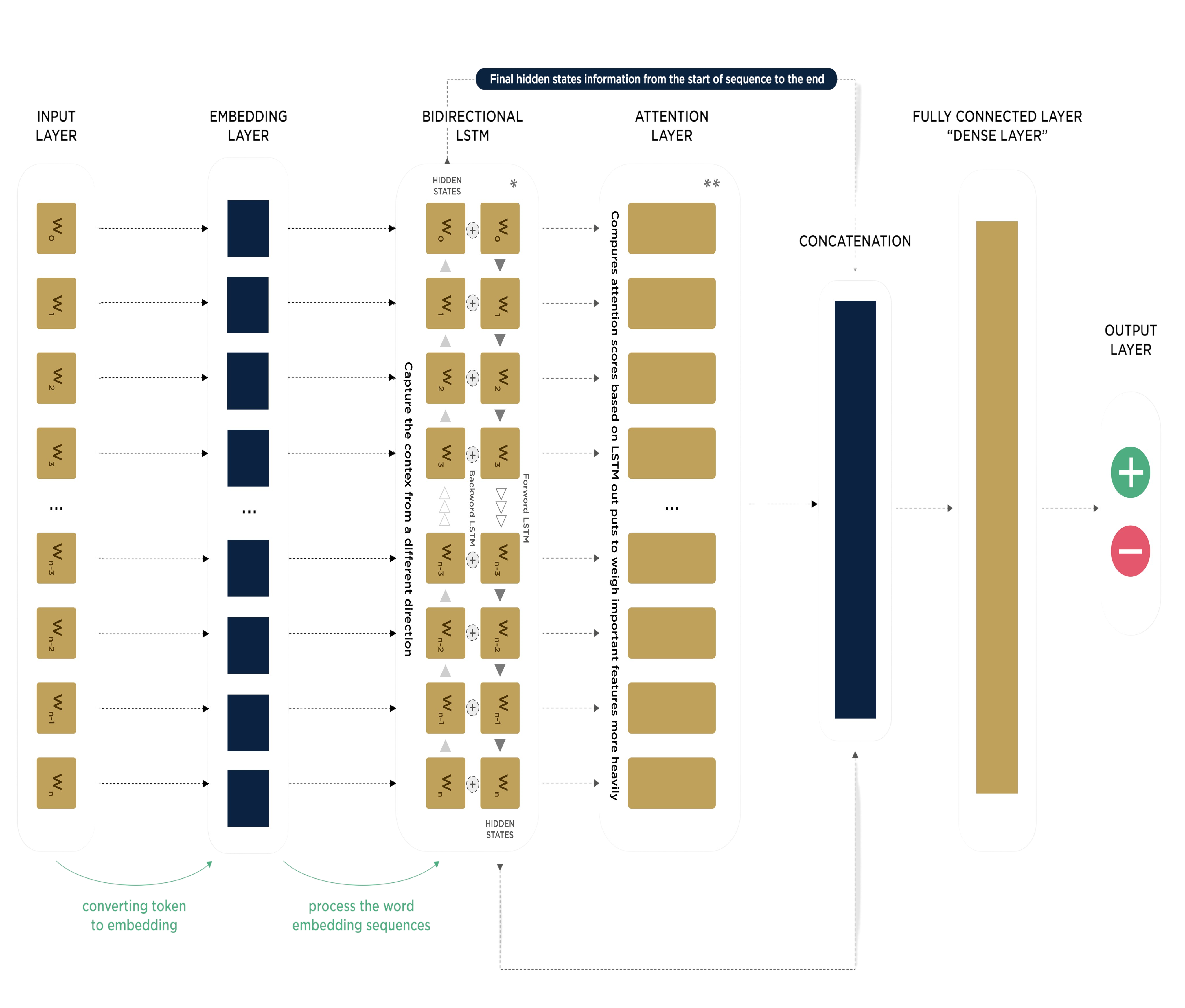}
    \caption{Architecture of the Proposed Sentiment Analysis Model. This figure illustrates our deep neural network design, highlighting the integration of the word embedding layer, bidirectional Long Short-Term Memory (LSTM) layer, attention mechanism, and output layer.  }
    \label{fig:1}
\end{figure}

Our proposed model leverages a deep neural network and primarily consists of four key components. From the bottom up in Figure \ref{fig:1}, these include the word embedding layer, Long Short-Term Memory (LSTM), attention mechanism, and the output layer. The embedding layer serves as the first layer, transforming each individual word of the input text data into dense vector representations. These vectors are used to capture the meanings and relationships between words in each piece of feedback. The embeddings are then processed by the LSTM layer, which handles the sequence of text in the feedback. Given that student feedback can be lengthy, we utilize LSTM, a type of recurrent neural network (RNN), to capture long-range dependencies within the data. The LSTM layer operates bidirectionally, enabling it to process feedback sentences both forwards and backwards. This approach helps the model achieve a comprehensive understanding of the feedback and the relationships between words. The output from the LSTM layer is then fed into the attention mechanism, which assigns varying weights to different parts of the input. The critical function of the attention layer is to focus on words that are more indicative of sentiment than others. The output layer, receiving input from the attention mechanism, generates raw scores (logits), which are passed through a sigmoid function to convert them to probabilities. This approach allows us to adjust the model's classification parameters without interpreting the output as a probability distribution, as would be the case with a softmax function in the output layer. Additionally, we round these probabilities to produce binary predictions (Positive or Negative), facilitating a comparison of the model's performance with the baseline model, which is a logistic regression.

\subsection{Model Evaluation Metrics}
\label{sec:metrics}
To evaluate our model against the baseline model, we have selected multiple evaluation metrics, including:

\subsubsection{Accuracy}
Accuracy measures the overall correctness of our model's predictions, representing the ratio of correct predictions to the total number of instances in the dataset. A higher accuracy indicates better model performance ~\cite{14}.

\subsubsection{Precision}
Precision is the proportion of true positive predictions to the total number of predicted positive instances. It measures the model's ability to identify true positives while avoiding false positives ~\cite{14}.

\subsubsection{Recall}
Recall, also known as sensitivity, represents the proportion of true positive instances and indicates our model's ability to capture all positive instances ~\cite{14}.

\subsubsection{F1 Score}
The F1 score is the mean of precision and recall, providing a balanced measurement between the two. A higher F1 score indicates a balanced and better-performing model ~\cite{14}.

\subsubsection{Area Under the ROC Curve AUC}
The Area Under the ROC curve provides an overall measurement of the model across all possible classification thresholds, for both positive and negative classes. The AUC ranges from 0 to 1, with higher values indicating better model performance ~\cite{14}.

\subsubsection{Confusion matrix}

The confusion matrix is as a binary classifier that evaluates the predictive accuracy of our model, containing True Negative (TN) and True Positive (TP), which imply that predictions are in alignment with actual results. However, the False Positive (FP) and False Negative (FN) indicates that predictions are not in alignment with actual results ~\cite{14}.

\section {RESULTS}

\subsection{Quantitative Analysis}

The Recurrent Neural Network (RNN) Model demonstrates superior performance compared to the baseline Logistic Regression (LR) Model across all metrics. Notably, the RNN model achieves a higher accuracy of 80\%, compared to the baseline's 77\%. This enhancement in the RNN model's performance is attributed to its ability to capture sequential word dependencies in sentences, which is crucial for understanding the contribution of each word to the overall meaning in sentiment feedback analysis. In terms of precision, the RNN model exhibits a superior performance with 83\%, compared to the logistic regression's 77\%. This outperformance indicates that the RNN model is more effective at minimizing false positives in predicting feedback cases. Furthermore, the RNN model excels in identifying true positive feedback cases, as reflected in its recall metric. Finally, the F1-score demonstrates that the RNN model outperforms the LR model by a margin of 7\%, indicating a more balanced improvement in both precision and recall.

\begin{table}[H]
\centering
\caption{Logistic Regression Model (Baseline Model) and Recurrent Neural Network Model Performance Evaluation}
\label{tab:classification_report}
\begin{tabular}{@{}lcccc@{}}
\toprule
Model Name & Accuracy & Precision & Recall & F1-Score \\
\midrule
Logistic Regression Model (Baseline) & 0.77 & 0.77 & 0.76 & 0.77 \\
Recurrent Neural Network Model & 0.80 & 0.83 & 0.85 & 0.84 \\
\bottomrule
\end{tabular}
\end{table}

The Recurrent Neural Network (RNN) model, with an ROC curve AUC (Area Under Curve) of 0.88, demonstrates proficient in distinguishing between positive and negative feedback cases. This performance is marginally superior to the Logistic Regression (LR) model, which presents an AUC of 0.86. While both models exhibit noticeable abilities, the RNN's higher AUC reflects its enhanced sensitivity and specificity in identifying true positives and true negatives. The RNN's architecture, which adept at capturing sequential and contextual feedback sentences in the dataset, likely contributes to this out performance over the LR model. The LR lacks to understating of sequential and contextual feedback sentence in the dataset, which inherent in RNNs.\cite{15}



\begin{figure}[H]
    \centering
    \begin{minipage}[t]{0.48\textwidth}
        \centering
        \includegraphics[width=\linewidth]{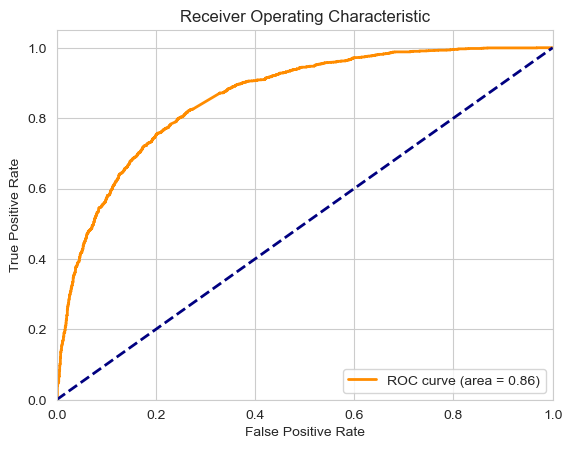}
        \caption{ROC curve displaying the performance of the sentiment analysis Logistic Regression (LR) model, with an Area Under the Curve (AUC) of 0.86.}
        \label{fig:roc_lr}
    \end{minipage}
    \hfill
    \begin{minipage}[t]{0.48\textwidth}
        \centering
        \includegraphics[width=\linewidth]{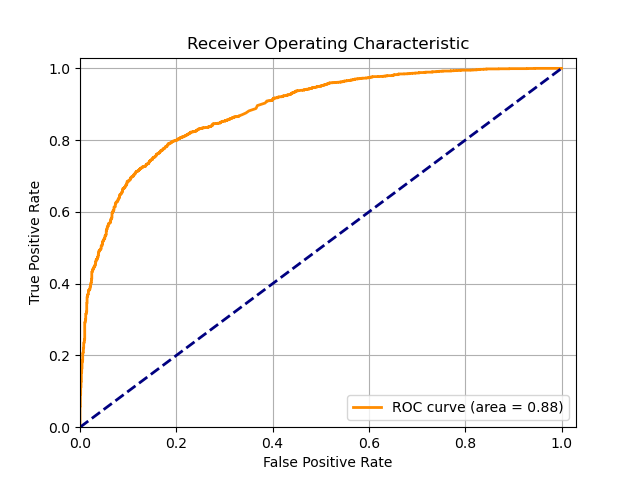}
        \caption{ROC curve displaying the performance of the sentiment analysis Recurrent Neural Network (RNN) model, with an Area Under the Curve (AUC) of 0.88.}
        \label{fig:roc_rnn}
    \end{minipage}
\end{figure}

The confusion matrix provides insights into the performance of The Recurrent Neural Network (RNN) model and Logistic Regression (LR) baseline model. The number of true positive predictions by the RNN model, where the model correctly identifies positive cases, is 2184, and the number of true negatives, where it correctly identifies negative cases, is 1009. Conversely, the model misclassified 422 cases as false positives and 381 as false negatives. In comparison, the LR baseline baseline model correctly identifies positive cases, is 1387, and the number of true negatives, where it correctly identifies negative cases, is 1718.  Conversely, the model misclassified 422 cases as false positives and 475 as false negatives. This breakdown highlights the RNN model's strengths in correctly classifying instances but also gives indications for the areas where the model may be prone to making errors.



\begin{figure}[H]
    \centering
    \begin{minipage}[t]{0.48\textwidth}
        \centering
        \includegraphics[width=\linewidth]{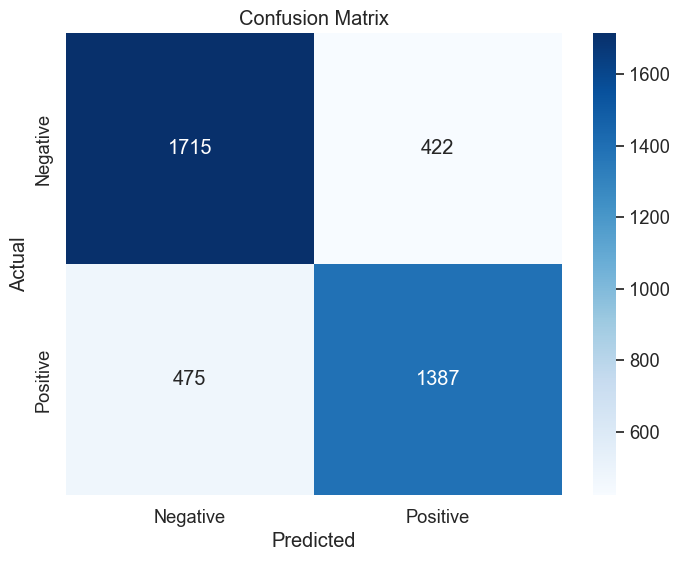}
        \caption{Confusion matrix illustrating the performance of the Logistic Regression (LR) baseline model on the test data.}
        \label{fig:cm_lr}
    \end{minipage}
    \hfill
    \begin{minipage}[t]{0.48\textwidth}
        \centering
        \includegraphics[width=\linewidth]{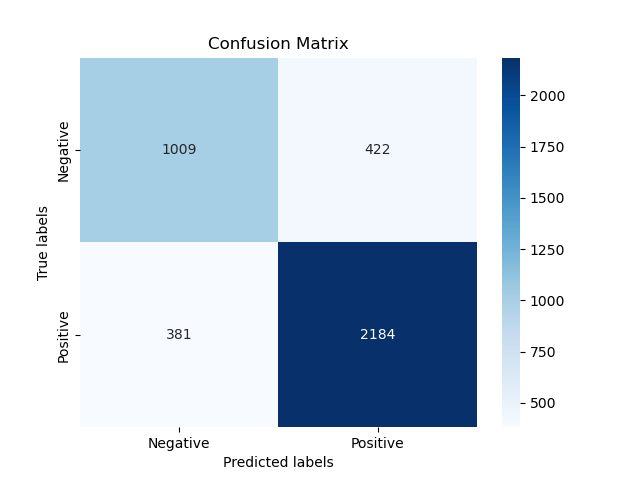}
        \caption{Confusion matrix illustrating the performance of the Recurrent Neural Network (RNN) model on the test data.}
        \label{fig:cm_rnn}
    \end{minipage}
\end{figure}

\subsection{Qualitative Analysis}

In the realm of model interpretability, our analysis delves into the decision-making processes of the Recurrent Neural Network (RNN). The RNN's ability to capture sequential dependencies \cite{15} aids in understanding how contextual information impacts predictions. In contrast, Logistic Regression (LR) offers straightforward interpretations by assigning clear weights to features, which simplifies the understanding of influencing factors \cite{16}. We evaluated various sentence variations in our qualitative analysis to comprehend how the RNN model interprets sentences compared to the LR model.

\begin{table}[H]
\centering
\caption{Sentence Variations for Analysis}
\label{tab:sentence_variations}
\begin{tabular}{|c|l|}
\hline
\textbf{Sentence No.} & \textbf{Sentence} \\ \hline
1 & The lecture was engaging and informative. \\ \hline
2 & Incredibly lecture but too long material. \\ \hline
3 & The lecture was conducted today. \\ \hline
4 & The lecture was extremely engaging and incredibly informative. \\ \hline
5 & The lecture was not engaging but informative. \\ \hline   
6 & The course material was engaging and informative. \\ \hline
7 & The lecture was informative but too long and tiring. \\ \hline
8 & The lecture was not engaging and informative. \\ \hline
\end{tabular}
\end{table}

\subsubsection{RNN Model observations}: The RNN model, with its ability to understand sequential dependencies \cite{15}, shows a higher positive probability for sentences with nuanced sentiments, such as "The lecture was extremely engaging and incredibly informative." Its advanced architecture allows it to capture subtle changes in sentiment, which is evident in sentences with mixed sentiments like "The lecture was informative but too long and tiring."


\subsubsection{LR Model Observations}: The LR model, on the other hand, tends to show more variation in probabilities for sentences with clearer sentiment demarcations. For instance, it assigns a higher negative probability to sentences like "The lecture was informative but too long and tiring." However, the sentence ended with negative word with starting with positive words, its interpretation seems less nuanced for complex sentences, possibly due to the lack of contextual and sequential data understanding \cite{16}.


\begin{figure}[H]
    \centering
    \begin{minipage}[t]{0.48\textwidth}
        \centering
        \includegraphics[width=\linewidth]{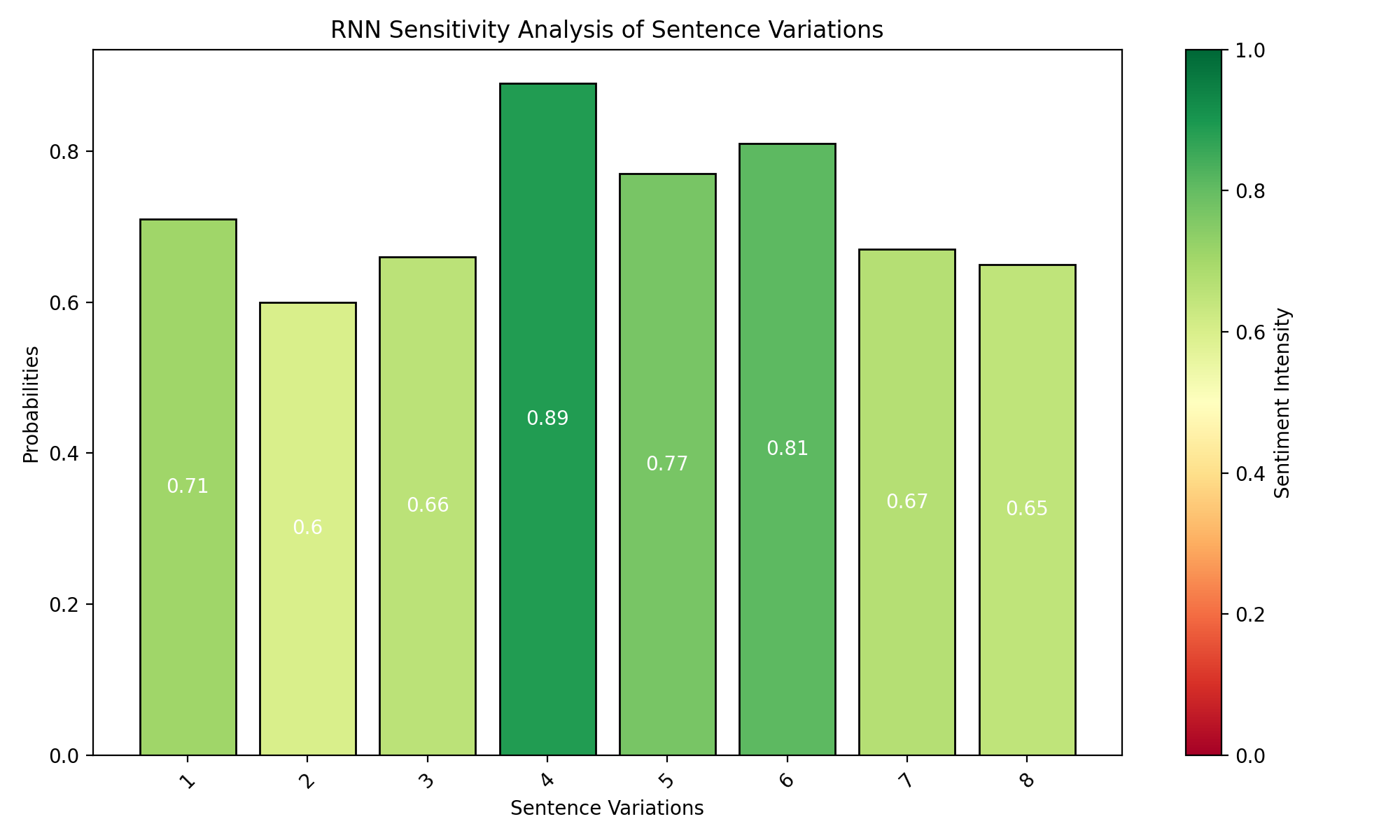}
        \caption{Sensitivity analysis of sentence variations using the Recurrent Neural Network (RNN) model, corresponding to sentences shown in Table~\ref{tab:sentence_variations}.}
        \label{fig:rnn_sa}
    \end{minipage}
    \hfill
    \begin{minipage}[t]{0.48\textwidth}
        \centering
        \includegraphics[width=\linewidth]{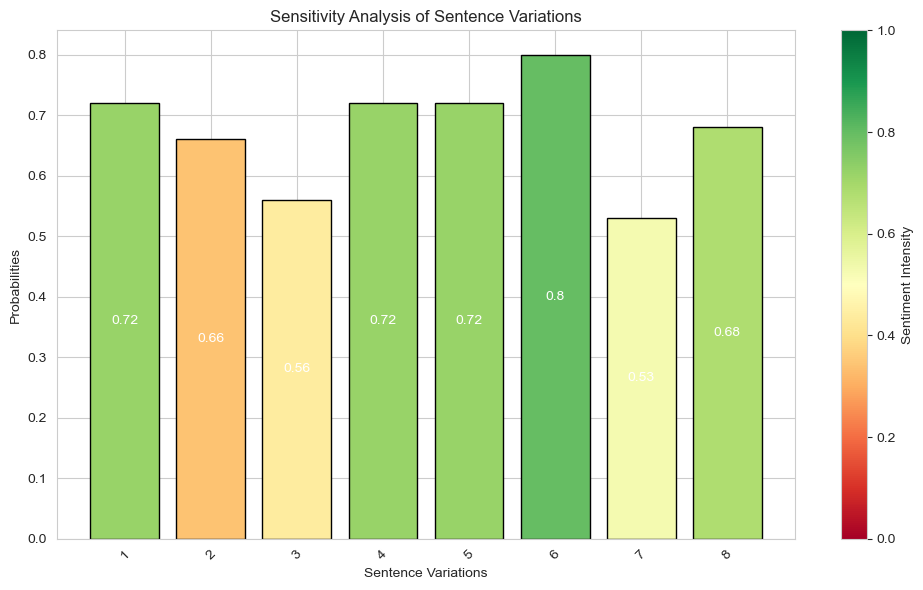}
        \caption{Sensitivity analysis of sentence variations using the Logistic Regression (LR) model, corresponding to sentences shown in Table~\ref{tab:sentence_variations}.}
        \label{fig:lr_sa}
    \end{minipage}
\end{figure}

\subsubsection{Comparative Analysis}: It is apparent that the RNN model excels in handling sentences with complex sentiment expressions, such as sentence 7 in Table \ref{tab:sentence_variations}, due to its sequence-based learning approach. The LR model, effective in clear-cut sentiment cases, may not as effectively capture context, particularly in sentences with mixed sentiments like sentence 7. Interestingly, both models failed to correctly classify sentence 8 as negative sentiment, despite the negative connotation implied by 'not' in sentence 8. Comparatively, in sentence 5, 'The lecture was not engaging but informative,' both models classified it as positive sentiment, with the RNN assigning a 65\% probability and LR a 68\% probability to positivity. This observation suggests that both models struggle with transferring negativity to subsequent words.

\section {Conclusion}
This research project successfully integrated a Recurrent Neural Network (RNN) model into a learning management system (LMS), offering an innovative approach to interpreting student feedback. The study demonstrates the RNN's effectiveness in capturing sentiment expressions, surpassing the baseline Logistic Regression model in various metrics. The RNN's advanced architecture is crucial for understanding sequential word dependencies, which enhances the perception of sentence sentiment. Integrating this model into an LMS provides educators with valuable insights into student feedback. While the RNN accurately labeled most sentiments, some inaccuracies highlight areas for future improvement. To enhance our model's capabilities, we propose incorporating transformer models, known for their superior performance in contextual and complex text understanding. This improvement would involve using contextual embedding \cite{17} to create a continuous representation for each word in the students' feedback, rather than building a global embedding. By employing contextual embedding, the model will better recognize the meaning of words in different contexts

\bibliographystyle{ACM-Reference-Format}
\bibliography{sample-base}


\end{document}